\documentclass{article}

\usepackage{graphicx}

\newcommand{\be}{\begin{equation}}
\newcommand{\ee}{\end{equation}}
\renewcommand{\k}[2]{\frac{#1}{#2}}

\def\p{\partial}

\def\s{\,\,\,\,}

\def\={\approx}
\def\a{\alpha}

\def\v{\lambda}
\def\ra{\rightarrow}
\newcommand{\pab}[2]{\frac{\p #1}{\p #2}}

\title{Density-Driven Compactional Flow in Porous Media}

\author{Xin-she Yang     \\
  Department of Applied Mathematics and Department of Fuel \& Energy \\
  University of Leeds, LEEDS LS2 9JT, UK }

\date{}

\begin{document}

\maketitle

\large

\begin{abstract}

In the mathematical modelling of compactional flow in porous media,
the constitutive relation is typically modelled in terms of a
nonlinear relationship between effective pressure and
porosity, and     compaction is essentially poroelastic.
However, at depths deeper  than $1$ km where pressure is high,
compaction becomes
more akin to a viscous one. Two mathematical models of compaction
in porous media are formulated and the noninear equations
are then solved numerically.  The essential features of numerical profiles of
poroelastic  and viscous compaction are thus compared with  asymptotic
solutions. Two distinguished styles of density-driven
compaction in fast and slow compacting sediments are analysed
and shown in this paper.


\noindent {\bf Citation detail:} 
X. S. Yang, Density-driven compactional flow in porous media, 
{\it Journal of Computational and Applied Mathematics},
{\bf 130}, 245-257 (2001).

\end{abstract}

\section{INTRODUCTION}

Density-driven compaction in porous media such as sediments
is an important process, which may occur in sedimentary basins where
hydrocarbons and oil are primarily formed. The modelling of such
density-driven flow is thus important in the oil industry as well
as in civil engineering. One particular problem which affects
drilling process is the occasional occurrence of abnormally
high pore fluid pressures, which, if encountered suddenly,
can cause drill hole collapse and consequent failure of the
drilling operation.   Therefore, an industrially important
objective is to predict overpressuring before drilling and
to identify its precursors during drilling.
An essential step to achieve such objectives
is the scientific understanding of their mechanisms and the
evolutionary history of post-depositional sediments such as shales.

Fine-grained sediments such as shales and sandstones are considered
to be the source rocks for much petroleum found in sandstones
and carbonates. At deposition, sediments such as shales and sands
typically have  porosities of order $0.5$ or $50 \%$. When sediments
are drilled at a depth, say 5000 m, porosities are typically
$0.05 \sim 0.2$ ($5 \% \sim 20 \%$)[1]. Thus an enormous
amount of water has escaped from the sediments during their deposition
and later evolution. Because of the fluid escape, the grain-to-grain
contact pressure must increase to support the overlying sediment
weight. Dynamical fluid escape depends lithologically on the permeability
behavior of the evolving sediments. As fluid escape proceeds, porosity
decreases, so permeability becomes smaller, leading to an ever-increasing
delay  in extracting the residual fluids. The addition of more overburden
sediments is then compensated for by an increase of excess pressure
in the retained fluids. Thus overpressure develops from such a
non-equilibrium compaction environment [2].
A rapidly accumulating basin is unable to expel pore fluids sufficiently
rapidly due to the weight of overburden rock. The development of
overpressuring retards compaction, resulting in a higher porosity, a higher
permeability and a higher thermal conductivity than are normal for a
given depth, which changes the structural and stratigraphic shaping of
sedimentary units and provides a potential for hydrocarbon migration.

Compaction is the process of volume reduction via
pore-water expulsion within sediments due to the increasing weight of
overburden load. The requirement of its occurrence is not only the
application of an overburden load but also the expulsion of pore water.
The extent of compaction is strongly influenced by burial history and
the lithology of sediments. The freshly deposited loosely packed
sediments tend to evolve, like an open system, towards a closely packed
grain framework during the initial stages of burial compaction and this is
accomplished by the processes of grain slippage, rotation, bending and
brittle fracturing. Such reorientation processes are collectively
referred to as  mechanical compaction,
which generally takes place in  the first 1 - 2 km of burial.
After this initial porosity loss,
further porosity reduction is accomplished by the process of
 chemical compaction such as pressure solution at grain contacts.
It is worth pointing out that  consolidation is a term often used
in geotechnical engineering and implies the reduction of pore space
by mechanical loading.
The fundamental understanding  of mechanical and
physico-chemical properties of  these rocks in the earth's crust
has important applications in petrology, sedimentology, soil mechanics,
oil and gas engineering and other geophysical research areas.
In spite of its geological importance,
the mechanism leading to pressure solution is still poorly understood[3].

The main aims in this paper are to determine and compare the essential
features of the  poroelastic and viscous compaction in a comprehensive way
and to understand these mechanisms by using new asymptotic
solutions and the comparison with full numerical simulations as well,
which will greatly extend the earlier work [2-4]. Another primary concern
of this paper is to try to formulate a new and more realistic
visco-poroelastic compaction relation.

\section{MATHEMATICAL MODEL}

For the convenience of investigating the effect of  compaction in porous
media  due to pure density differences, we will assume the basic model
of compaction is rather analogous to the process of soil consolidation.
The porous  media act as a compressible porous matrix, so that mass
conservation of pore fluid together with  Darcy's law leads to the
1-D model equations of the general type [3,4]. Let $t$ be time and
$z$ be the space co-ordinate directing upwards, the governing equations
can be written as
\begin{equation}
\pab{[\rho_s (1-\phi)]}{t} +\pab{}{z}[\rho_s (1-\phi) u^{s} ]=0,
\s ({\rm solid \,\, phase})
\label{VC:MASS}
\end{equation}
\begin{equation}
\pab{(\rho_l \phi)}{t}+ \pab{(\rho_l \phi  u^{l})}{z}=0,
\s ({\rm liquid \,\, phase})
\end{equation}
\begin{equation}
\phi (u^{l}-u^{s})
=\frac{k(\phi)}{\mu}[G \pab{p_e}{z}-(\rho_s-\rho_l)(1-\phi) g ],
\s ({\rm Darcy's law})
\end{equation}
where $\phi$ is the porosity of the pores saturated with water.
$u^{l}$ and $u^{s}$ are the velocities of fluid and
solid matrix, $k$ and $\mu$ are the matrix permeability and
the liquid viscosity,   $\rho_{l} $ and $\rho_{s}$ are the
densities of fluid and solid matrix,  $p_{e}$ is the effective pressure,
$G$ is a constant of the properties in porous media, and $g$ is the
gravitational acceleration. In addition, a  compaction relation
is needed to complete this  model [4,5]. By assuming the densities
$\rho_s$ and $\rho_l$ are constants, we can see that only the density
difference $\rho_s-\rho_l$ is important to the flow evolution. Thus,
the compactional flow is essentially density-driven flow in porous
media.

\subsection{Poroelasticity and Viscous Compaction}

Compaction relation is a relationship between effective pressure $p_e$ and
strain rate $\dot e=\pab{u^s}{z}$ or porosity $\phi$ [6]. The common approach
in soil mechanics and sediment compaction is to model this generally nonlinear
behaviour as  poroelastic, that is to say, a relationship of Athy's law type
$p_e = p_e (\phi)$, which is derived from fitting the real data of sediments.
Athy's poroelasticity law is also a simplified form of Critical State Theory.
A common relation representing the poroelasticity is
\begin{equation}
\frac{D p_e}{D t}=-K_s \pab{u^s}{z},   \,\,\,\,\, \frac{D}{D t}=
\frac{\partial}{\partial t} + u^s \pab{}{z}, \label{equok}
\end{equation}
where $K_s$ is a modulus of   sediment compression.
As $\rho_s$ is a constant and can thus be eliminated by multiplying equation
(\ref{VC:MASS}) by $1/\rho_s$, and we  get
\begin{equation}
\pab{(1-\phi)}{t}+u^s \pab{(1-\phi)}{z}=-(1-\phi) \pab{u^s}{z}, \s
{\rm or} \s
\frac{1}{1-\phi} \frac{D (1-\phi)}{D t} = - \pab{u^s}{z},
\end{equation}
combining with the previous equation (\ref{equok}), we have
\begin{equation}
p_e = p_e (\phi),
\end{equation}
which is the Athy's law for poroelasticity. However, this
poroelastic compaction law is  only valid for the  compaction in porous media
in the upper and shallow region, where compaction  occurs due to the pure
mechanical movements such as  grain sliding and packing rearrangement. In the
more deeper region, mechanical compaction is gradually replaced by the
chemical compaction due to stress-enhanced flow along the grain boundary
from the grain contact areas to the free pore, where pressure is essentially
pore pressure. A typical process of such chemical compaction in sediment
is pressure solution whose rheological behavior is usually viscous, so that
it sometimes called viscous pressure solution or viscous creep.

The mathematical formulation for viscous compaction is to derive a relation
between creep rate $\dot e$ and effective stress $\sigma_e$. Rutter's
creep relation is widely used [7,8]
\begin{equation}
\dot e=\frac{A_{k} c_{0} \, w D_{gb} }{\rho_{s} \bar d^{3}}
\sigma_e, \label{CREEP-1}
\end{equation}
where $\sigma_e$ is the effective normal stress across the grain contacts,
$A_{k}$ is a constant, $c_{0}$ is the equilibrium concentration
(of quartz) in pore fluid,  $\rho, \, \bar d$ are the density
and (averaged) grain diameter (of quartz). $D_{gb}$ is the diffusivity
of the solute in water along grain boundaries with a thickness $w$.
Note that $\sigma_e=-G p_e$ and $\dot e=\pab{u^{s}}{z}$.
With this, (\ref{CREEP-1}) becomes the following  compaction law
\begin{equation}
p_{e}=-\xi \nabla . {\bf u}^{s},   \s \xi=\k{ G\rho_{s} \bar d^{3}}{A_{k} c_{0} \, w D_{gb} }.
\label{VC:CREEPNEW}
\end{equation}

More generally speaking, $\xi$ is also a function of porosity $\phi$.
The compaction law is analogous to the viscous compaction laws
used in studies of magma transport in the Earth's mantle [9,10].

\subsection{Boundary conditions}

The boundary conditions for the governing equations
are as follows. The bottom boundary at $z=0$ is assumed to be
impermeable
\begin{equation}
u^{s}=u^{l}=0,
\end{equation}
and a top condition at $z=h$ is kinetic
\begin{equation}
\dot{h}=\dot{m}_{s}+u^{s}, \label{eq:ht}
\end{equation}
where $\dot{m}_{s}$ is the sedimentation rate at $z=h$. Also at $z=h$,
\begin{equation}
\phi=\phi_{0}, \s p_e=p_0,
\end{equation}
where $p_0$ is the applied effective pressure at the top of the
porous media, and $\phi_{0}$ is the initial porosity.

\section{Non-dimensionalization}

If a length-scale $d$ is a typical length [8] defined by
\be
d=\{\k{\xi \dot m_s G}{(\rho_s-\rho_l) g}\}^{\k{1}{2}},
\ee
and the effective pressure is
scaled in the following way
\begin{equation}
p=\k{G (p_{e}-p_0)}{(\rho_{s}-\rho_{l}) g d},
\end{equation}
so that $p=O(1)$. Meanwhile, we  scale $z$ with $d$, $u^{s}$ with
${\dot m}_{s}$, time $t$ with $ d/{\dot m}_{s}$, permeability $k$ with
$k_{0}$. By writing $k(\phi)=k_{0} k^*$, $z=d z^*$, ..., and dropping
the asterisks, we thus have
\begin{equation}
-\pab{\phi}{t} +\pab{}{z}[(1-\phi) u^{s}]=0,
\label{VC:MASS-1}
\end{equation}
\begin{equation}
\pab{\phi}{t}+ \pab{(\phi u^{l})}{z}=0, \label{PHI:U}
\end{equation}
\begin{equation}
\phi (u^{l}-u^{s})=\v k(\phi) [\pab{p}{z}-(1-\phi) ].
\end{equation}
The poroelastic relation becomes
\be
p=p(\phi)
\ee
and the viscous relation is
\be
p=-\pab{u^s}{z}.
\ee
where
\begin{equation}
\lambda=\frac{k_{0} (\rho_{s}-\rho_{l}) g} {\mu {\dot m}_{s}}.
\ee
Adding (\ref{VC:MASS-1}) and (\ref{PHI:U}) together and integrating
from the bottom, we have
\be
u^s=-\phi (u^l-u^s)=-u,
\ee
where $u=\phi (u^l-u^s)$ is the Darcy flow velocity. Now we have
\begin{equation}
\pab{\phi}{t} +\pab{}{z}[(1-\phi) u]=0,
\end{equation}
\begin{equation}
u=-\v k(\phi) [\pab{p}{z}-(1-\phi) ].
\end{equation}
The constitutive relation for permeability $k(\phi)$ is nonlinear [11], and
its typical form  is
\be
k(\phi)=(\k{\phi}{\phi_0})^m, \s m=8.
\ee
Different formulations of compaction relation may lead to different
compaction models.  One way is to use a relationship between effective
pressure $p$ and matrix velocity $u^s$ (or ${\bf u}^s$ in 3-D form)
as given in (\ref{VC:CREEPNEW}).
However, a more common way is to write a relation between $p$ and
porosity $\phi$. Formulating the compaction relation in this way,
we  have\\
\noindent {\it Poroelastic Model}:
\begin{equation}
\pab{\phi}{t} =\v \pab{}{z} \{ (1-\phi) (\k{\phi}{\phi_0})^m
[\pab{p}{z}-(1-\phi)] \}.
\label{equ-100}
\end{equation}
\be
p=\k{1}{\a} [\ln \k{\phi_0}{\phi}-(\phi_0-\phi)],
\label{equ-200}
\ee
which is a relation of Athy-type. $\a=O(1)$ is usually called the compaction
or consolidation coefficient. The boundary conditions are
\be
\pab{p}{z}-(1-\phi)=0, \s  {\rm at } \s z=0,
\label{bbb-1-0}
\ee
\be
\phi=\phi_{0},  \s
\dot h=\dot m(t) + \v (\k{\phi}{\phi_0})^m [\pab{p}{z}-(1-\phi)]
\s {\rm at} \s z=h(t).
\label{bbb-1-1}
\end{equation}

\noindent {\it Viscous Model}:
\begin{equation}
\pab{\phi}{t} = \v \pab{}{z}\{ (1-\phi)
(\k{\phi}{\phi_0})^m [\pab{p}{z}-(1-\phi)] \},
\label{equ-300}
\end{equation}
\begin{equation}
p=\v \pab{}{z} \{ (\k{\phi}{\phi_0})^m [\pab{p}{z}-(1-\phi) ] \},
\label{equ-400}
\end{equation}
The boundary conditions are
\be
\pab{p}{z}-(1-\phi)=0, \s {\rm at } \s z=0,
\label{bbb-2-0}
\ee
\be
\phi=\phi_{0},  \s
\dot h=\dot m(t) +\v (\k{\phi}{\phi_0})^m [\pab{p}{z}-(1-\phi) ] \,\,
\s {\rm at} \s z=h(t).
\label{bbb-2-1}
\end{equation}
where $\dot m (t)=O(1)$ is a prescribed function of time, which can be taken
to be one for constant sedimentation on top of the porous media.  Obviously,
$\dot m=0$ if there is no further sedimentation and no increasing  loading
on top of the porous media.

It is useful for the understanding of the solutions to get an estimate
for $\lambda$ by using values taken from observations and earlier
work [1, 4, 11]. By using the
typical values of $\rho_{l} \sim 10^{3} \, {\rm kg\, m}^{-3},\,
\rho_{s} \sim 2.5 \times 10^{3} \, {\rm kg\, m}^{-3},\,$
$ k_{0} \sim 10^{-15} -\!\!- 10^{-20}\, {\rm m}^{2}, \, \mu \sim 10^{-3}\,
{\rm N\,s\,
m}^{2}, \, \xi  \sim 1 \times 10^{21}$ N s $ {\rm m}^{-2}, $
$\dot m_{s}  \sim 300\, {\rm m\,\, Ma}^{-1}=1 \times 10^{-11}\,
{\rm m\,\, s}^{-1},\, g \=10 {\rm m \,s}^{-2}, \, G \=1$;
then $\v \= 0.01 -\!- 1000$ and $d \= 1000$ m. Therefore, $\lambda=1$
defines a transition between the slow  compaction ($\lambda << 1$) and fast
compaction ($\lambda >> 1$). The parameter $\lambda$ , which is the ratio
between the permeability and the sedimentation rate, governs the
evolution of the pore pressure and porosity in sedimentary basins.  High
sedimentation rate may gives rise to excess pressures even in the basins
with moderate permeability.

\section{Numerical Simulations and Asymptotic  Analysis}

\subsection{Numerical Method}

In order to solve the highly coupled non-linear equations, an implicit
numerical difference method is used [12]. Substituting the expression for
effective pressure $p$ into the $\phi$ equation, the essential
equation for porosity $\phi$ becomes the standard non-linear
parabolic form
\begin{equation}
\phi_{t}=F(z,t,\phi) \phi_{zz}+g(z,t,\phi, \phi_{z}).
\end{equation}
The first stage gives $\phi^{n+1/2}$ as a solution of the following equation
\[
\frac{2}{\Delta
t}(\phi^{n+1/2}_{i}-\phi^{n}_{i})
=(\frac{1}{\Delta
z^2})F(z_{i},t^{n+1/2},\phi^{n}_{i})\delta^{2}_{z} \phi^{n+1/2}_{i} \]
\begin{equation}
+g(z_{i},t^{n+1/2},\phi^{n}_{i},\frac{1}{\Delta
z}\delta_{z} \phi^{n}_{i}),
\end{equation}
where $\delta^{2}_{z}\phi_{i}=(\phi_{i+1}-2\phi_{i}+\phi_{i-1})$ and
$\delta_{z}\phi_{i}=(1/2)(\phi_{i+1}-\phi_{i-1})$. $\Delta t$ and $\Delta z$
are the time and space increments after discretisation, respectively.
The second stage
gives $\phi^{n+1}_{i}$ as a solution of the following equation
\[ \frac{1}{\Delta
t}(\phi^{n+1}_{i}-\phi^{n}_{i})=(\frac{1}{2(\Delta
z)^2})F(z_{i},t^{n+1/2},\phi^{n+1/2}_{i})\delta^{2}_{z}(\phi^{n+1}_{i}
+\phi^{n}_{i})
\]
\begin{equation}
+g(z_{i},t^{n+1/2},\frac{1}{\Delta
z}\delta_{z} \phi^{n+1/2}_{i}).
\end{equation}
The convergence is second-order in space for this method, and
$O(\Delta t)^{2-\epsilon}$ in time, where $\epsilon $ is a small
number less than 1/2.

The computational convergence of the calculation of this method has been
tested by 1)  changing the number of grid per unit ($1/\Delta z$)
from 5 to 1000 in space and $1/\Delta t$ from 10
to 5000 in time, and by 2) comparing with the results of asymptotic results.
The changes of grid intervals all result in the same converged results
which conform well to the asymptotic solutions. This shows that this method is
robust for the solution of the equations encountered in our problems.

\subsection{Numerical Results}

We used a normalized grid  by employing the rescaled height
variable $Z=z/h(t)$ in a fixed domain, which will make it easy
to compare the results of different times with different values
of dimensionless parameters in a fixed frame. This transformation
maps the basement of the basin to $Z=0$ and the basin top to $Z=1$.
The calculations were mainly implemented for the time evolutions
in the range of $t=0.5 \sim 10$ corresponding to the real time range
$1.5 \sim 30 $ million years and the real range in thickness
is $0.5 \ {\rm km} \sim 10 \ {\rm km}$ which is the one of main interest
in the petroleum industry. In addition, the timescale can be chosen
in such a way that $t=0.5 \sim 10$ corresponding to  the real time in
the order of $15$ days to $20$ years with a real thickness
from $5 \sim 1500$ m in civil engineering.
Numerical results are briefly
presented and explained  below. The comparison with the asymptotic
solutions for equilibrium state will be made in the next section.

Figure 1 shows the poroelastic compaction  profile  of porosity
$\phi$ versus the  rescaled  height $Z$ at different times
$t=1,2,3,5,8$. The value of $\v=100$ has been used in the calculations.
We can see that porosity decreases quite dramatically at the top,
and profile is nearly exponential versus the rescaled depth $1-Z$.

Figure 2 provides the viscous compaction profile of porosity
versus the rescaled height. All the  other  parameters are the same. The
only difference from that of Figure 1      is   that the compaction
relation is now viscous. Comparing with the profile in  Figure 1,
it is   clearly seen that  porosity changes less slowly than that
in the poroelastic case. The profile now is more or less parabolic.
Although these two figures are quite different in the top region,
there are still some similarity in the lower region, where the
porosity decrease very slowly due to the fact that permeability $k(\phi)
=(\phi/\phi_0)^m$ is getting virtually very small as $\phi<\phi_0$
and $m=8$, which will in turn constrain the density-driven flow
through the porous media, and thus consequently slow down the compaction
process.

To understand these phenomena and to verify these numerical results,
it would be very helpful if we can find some analytical solutions to
be compared with. However, it is very difficulty to get general
solutions for poroelastic compaction equations (\ref{equ-100}) and
(\ref{equ-200})  or viscous compaction  equations  (\ref{equ-300})
and (\ref{equ-400})   because these equations are nonlinear with a
moving boundary $h(t)$.  Nevertheless,  it is still possible and
very helpful to  find out the equilibrium state and compare
with the full numerical solutions.

\subsection{Equilibrium State}

To find out the solutions for  the equilibrium state, we must solve
a nonlinear or a pair of nonlinear ordinary differential equations
whose solution can usually implicitly be written in the quadrature  form.
In order to plot out and see the insight of the mechanism, we also
need to solve these ODEs numerically although  the solution procedure is
straightforward. However, it is practical    to get the asymptotic
solutions in the explicit form in the     following cases.

\subsubsection{Poroelastic Compaction}

For the poroelastic compaction, the equations for equilibrium state become
\be
\v \pab{}{z} \{ (1-\phi)
(\k{\phi}{\phi_0})^m [\pab{p}{z}-(1-\phi)] \} =0.
\ee
Substituting the expression for $p$ and integrating the above equation once
together with the top boundary condition (\ref{bbb-1-0}) gives
\be
\v (1-\phi)^2 (\k{\phi}{\phi_0})^m [\k{1}{\a \phi} \pab{\phi}{z}-1]
=(\dot m-\dot h)(1-\phi_0),
\ee
where we have assumed that $\dot m(t)=1$ and $\dot h=const$.  The solution of
this equation can be written in a quadrature although it is nonlinear.

Since $\v=0.01 -\!\! 1000$, we can expect that two distinguished limits
$\v \ra 0$ and $\v \ra \infty$ will have very different features.
For $\v \ra 0$, we have
\be
\dot h=\dot m, \s \phi \= \phi_0,
\ee
which means that porosity does not change and no compaction occur. This
corresponds to the case of very fast sedimentation or the density difference
$\Delta \rho=\rho_s-\rho_l \ra 0$.
On the other hand, as $\v \ra \infty$, we have
\be
[\k{1}{\a \phi} \pab{\phi}{z}-1] \= 0,
\ee
its solution with the top boundary condition can be straightforwardly
written as
\be
\phi=\phi_0 e^{-\a (h-z)},  \label{equ-500}
\ee
which is essentially the Athy's profile derived from real field data in
sedimentary basins. Clearly, if $\a \ra 0$ (very slow consolidation),
$\phi \= \phi_0$, which means that porosity changes also very slow.
If $\a \ra \infty$ (very quick consolidation), $\phi \ra 0$ for $h-z>1/\a$,
which implies that compaction proceeds so fast that the porosity  is
virtually zero everywhere except in a thin boundary region at the top.
The thickness of the top boundary layer is approximately $1/\a$, which
is usually $O(1)$. However, the solution (\ref{equ-500}) also satisfies the
bottom boundary condition $\pab{\phi}{z}-\a \phi=0$ at $z=0$, which means
that this solution is a uniformly valid solution for steady state.

\subsubsection{Viscous Compaction}

For the viscous compaction, the equilibrium state is governed by
\[ \v \pab{}{z}\{ (1-\phi)
(\k{\phi}{\phi_0})^m [\pab{p}{z}-(1-\phi)] \}=0, \]
\begin{equation}
p=\v \pab{}{z} \{ (\k{\phi}{\phi_0})^m [\pab{p}{z}-(1-\phi) ] \},
\ee
The integration of the first equation together with the top  boundary
condition leads to
\be
p=\pab{}{z}[\k{(\dot m-\dot h)(1-\phi_0)}{1-\phi}],
\ee
and
\be
\k{(\dot m-\dot h)(1-\phi_0)}{1-\phi}
=\v  (\k{\phi}{\phi_0})^m  [ (\dot m-\dot h)(1-\phi_0)
\k{\p^2}{\p z^2} (\k{1}{1-\phi})-(1-\phi) ],
\ee
whose general solution can also be written in a quadrature. However,
two  distinguished limits are more interesting. Clearly, if $\v \ra 0$,
we have
\be
\dot h=\dot m, \s \phi=\phi_0,
\ee
which is the case of no compaction as discussed in the case of poroelastic
compaction. Meanwhile, if  $\v \ra \infty$, we have
\be
(\dot m-\dot h)(1-\phi_0)
\k{\p^2}{\p z^2}(\k{1}{1-\phi})-(1-\phi)=0,
\ee
which can be rewritten as
\be
(\dot m-\dot h)(1-\phi_0) \psi''-\k{1}{\psi}=0, \s \psi=\k{1}{1-\phi}.
\ee
By using $\psi''=\psi d \psi'/d \psi$ and integrating from $h$ to $z$,
we have
\be
\k{(\dot m-\dot h)(1-\phi_0)}{2} (\psi')^2=\ln \k{\psi}{\psi_0},
\s \psi_0=\k{1}{1-\phi_0}.
\ee
Further integration leads to
\be
i [{\rm erf}{\k{i}{1-\phi}}-{\rm erf}{\k{i}{1-\phi_0}}]=\sqrt{\k{2(1-\phi_0)}
{\pi (\dot m-\dot h)}} (h-z).
\label{equ-500-500}
\ee
The comparison of poroelastic solution (\ref{equ-500}) and viscous solution
(\ref{equ-500-500}) with the numerical results is shown in Figure 3 in
the top region where the compaction profile is nearly at equilibrium state
for  $\v=1000$ and $t=10$. The clearly agreement verifies the numerical
method and the asymptotic   solution procedure.

\section{Discussions}

Conventional studies of compaction in porous media have
focused on the separate   features of poroelastic and
viscous compaction.  The  novelty  of this paper is
to compare and find out distinguished features of these
two different compaction styles.

Based on the  pseudo-steady state approximations, the
model equations  of compaction can be  simply  written
in dimensionless form as a mass conservation and Darcy's law.
A constitutive compaction relation is needed to complete this model.
In the case of poro-elastic compaction, we use an Athy-type relation
$\tilde p=\tilde p(\phi)$; while in the case of viscous compaction
due to pressure solution creep only, we choose
$\tilde p=-\k{\partial u^{s}}{\partial z}$.
These two different relations result in two quite
different behaviours of porosity evolution.
In the simpler poro-elastic case, we have
a single non-linear diffusion equation for porosity $\phi$.

The analysis showed that the limit $\lambda \ra 0$ (very
slow compaction) can be simply analysed by means of a boundary
layer analysis at the sediment base. The more interesting
mathematical case is when $\lambda>>1$ (fast compaction).
For sufficiently small times, the porosity profile is exponential
with depth, corresponding to an equilibrium (very long time)
profile. However, because of the large exponent $m$ in the
permeability law $\tilde k=(\phi/\phi_{0})^{m}$, we find that
even if $\lambda>>1$, the product $\lambda \tilde k$ may
become small at sufficiently large depths. In this case, the
porosity profile consists of an upper part
near the surface where $\lambda \tilde k>>1$ and the equilibrium is
attained, and a lower part where $\lambda \tilde k<<1$, and the
porosity is higher than equilibrium which appears to
correspond accurately to numerical computations. For the case of
viscous compaction,    porosity reduction occurs
throughout the basin, and the basic equilibrium solution
which applies near the surface is a near parabolic profile of porosity.
The differences in these two profiles are very distinguished.

From the solution (\ref{equ-500}) for poroelastic compaction at
equilibrium state,
we see that $\phi \ll \phi_0$ when $\a (h-z)=O(1)$ or $(h-z)=O(1/\a)$, that is
to say,  the solution is significant in a region shallower than
\be
\Pi_p \=\k{d}{\a},
\ee
which corresponds to a depth of 1000 m when $d \=1000$ m and $\a=1.0$.
On the other hand, the viscous solution (\ref{equ-500-500}) only becomes
significant when $\sqrt{\pi (\dot m-\dot h)/2(1-\phi_0)}=O(1)$, or in
the region of depths $h-z$ greater than
\be
\Pi_v \=d \sqrt{\k{\pi (\dot m-\dot h)}{2 (1-\phi_0)} },
\ee
which is equivalent to a depth of 970 m with values of $\phi_0=0.5$,
$\dot m-\dot h=0.3$ and $d=1000$ m. Therefore, we can generally  anticipate
that the poroelastic compaction is dominant in the shallow region
from the surface to a depth of 1 km. At depths greater  than 1 km,
the pressure is high enough,
pressure solution mechanism becomes significant and thus
compaction is essential viscous.  Naturally, there exists a region
of depths near 1km  where both mechanism becomes  important, and
an obvious extension is to include both models in a more realistic
model.    From the poroelastic constitutive relation (\ref{equok})
and viscous  relation (\ref{VC:CREEPNEW}), we can  formulate
a generalised viscous-poroelastic compaction model of Maxwell type
\be
{\bf \nabla . u^s}=-\k{1}{K_s} \k{D p_e}{D t}-\k{1}{\xi} p_e.
\ee
Subsequently, we would expect a visco-poroelastic porous medium
and thus some care is needed to ensure the
resulting model involving material derivatives is frame invariant.
Fortunately, this frame invariance is alway true in the present 1-D
formulation.    Incorporation of these extension and other processes
such as convection and 3-D density-driven flow will form the substance
of future work. \\

{\bf Acknowledgements.}  The author wishes to thank the anonymous referees
for their very helpful comments and very instructive suggestions.

\section*{References}

\begin{description}
{\small 
\item[1]  I. Lerche, {\it Basin Analysis: Quantitative Methods} (
Academic Press, San Diego, California, 1990).

\item[2]  R. E. Gibson, G. L. England and M. J. L. Hussey, The Theory of
One-dimensional Consolidation of Saturated Clays, I. Finite Non-linear
Consolidation of Thin Homogeneous Layers, {\em Can. Geotech. J.}, {\bf
17}(1967)261-273.

\item[3]   D. M. Audet and A. C. Fowler, A Mathematical Model for
Compaction in Sedimentary Basins, {\it Geophys. Jour. Int.}, 1992,
{\bf 110} (1992) 577-590.

\item[4]   A. C. Fowler and X. S. Yang, Fast and Slow Compaction
           in Sedimentary Basins, {\it SIAM Jour. Appl. Math.}, {\bf 59}(1998)
           365-385.

\item[5]   J. Bear and Y. Bachmat, {\em Introduction to Modeling of
Transport Phenomena in Porous Media} (Kluwer Academic, London, 1990).

\item[6]   M. A. Biot, M.A., General Theory of Three-dimensional
Consolidation, {\em J. Appl. Phys.}, {\bf 12} (1941) 155-164.

\item[7]   E. H. Rutter, Pressure Solution in Nature, Theory and
Experiment, {\em J. Geol. Soc. London}, {\bf 140} (1976) 725-740.

\item[8]   X. S. Yang,  Mathematical Modelling of Compaction and Diagenesis
           in Sedimentary Basins, D.Phil Thesis, Oxford University,
           1997.

\item[9]   D P McKenzie,  The generation and compaction of
           partial melts, {\em J. Petrol.}, {\bf 25} (1984) 713-765.

\item[10]  A C Fowler, A compaction model for melt transport in the
           Earth's asthenosphere. Part I: the basic model, in {\em Magma
           Transport and Storage}, ed. Ryan, M.P., John Wiley (1990) 3-14.

\item[11]   Smith, J.E., The dynamics of shale compaction and evolution in
           pore-fluid pressures, {Math. Geol.}, 1971, {v.3}, 239-263.

\item[12]  P.C. Meek and J. Norbury,  Two-Stage, Two Level Finite
           Difference Schemes for Non-linear Parabolic Equations,
           {\it IMA J. Num. Anal.}, {\bf 2}(1982) 335-356.
}
\end{description}

\def\fig#1#2{\begin{figure} \centerline{\includegraphics[width=4.5in,height=3in]{#1}} \caption{#2} \end{figure}}

\fig{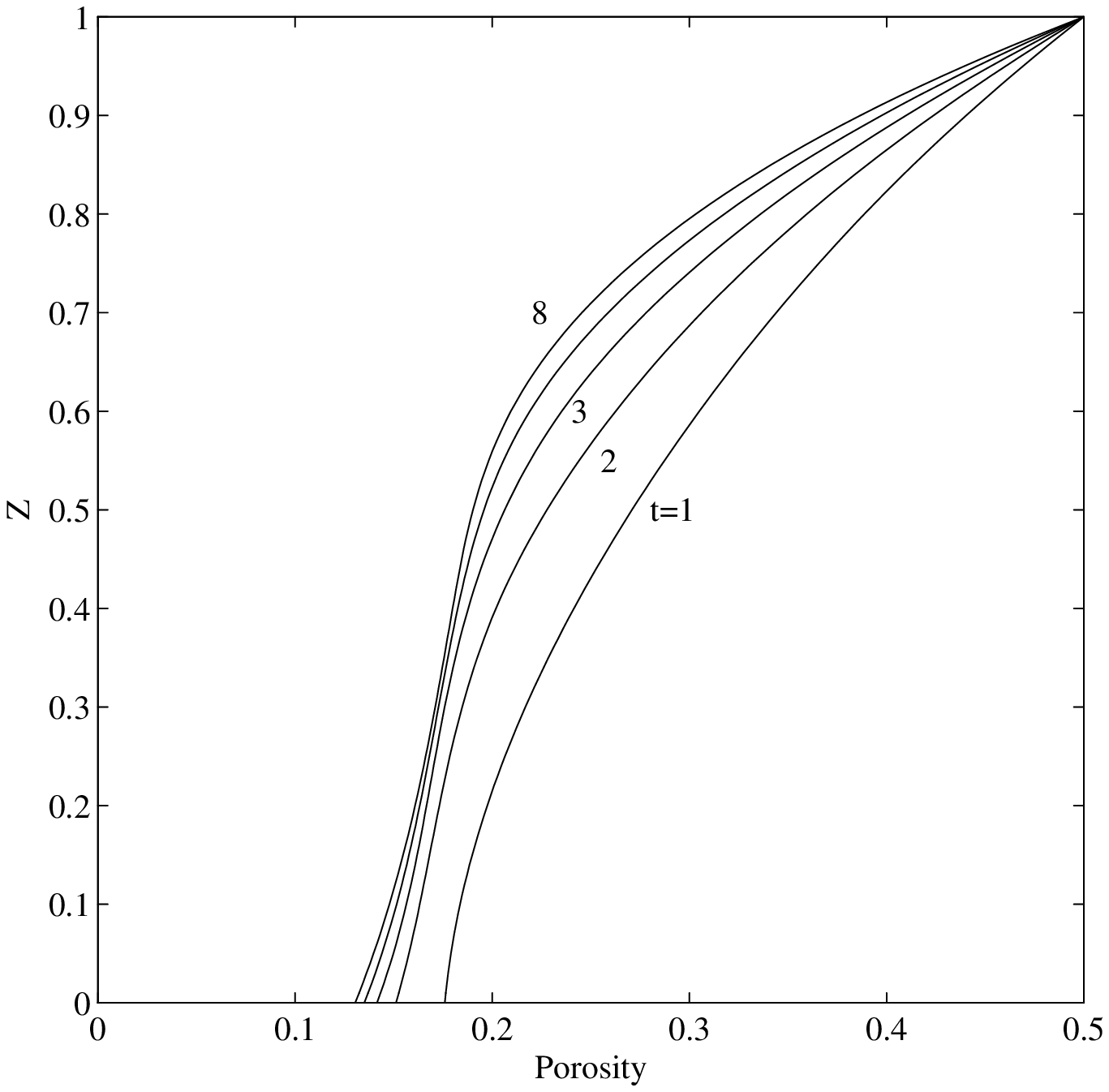}{ Poroelastic compaction  profile of porosity versus
         rescaled height $Z=z/h(t)$ at different times $t=1,2,3,5,8$
         for $\v=100$. Athy's law between porosity and effective pressure
         is  used. Porosity decreases  essentially exponentially
         in the top region. }

\fig{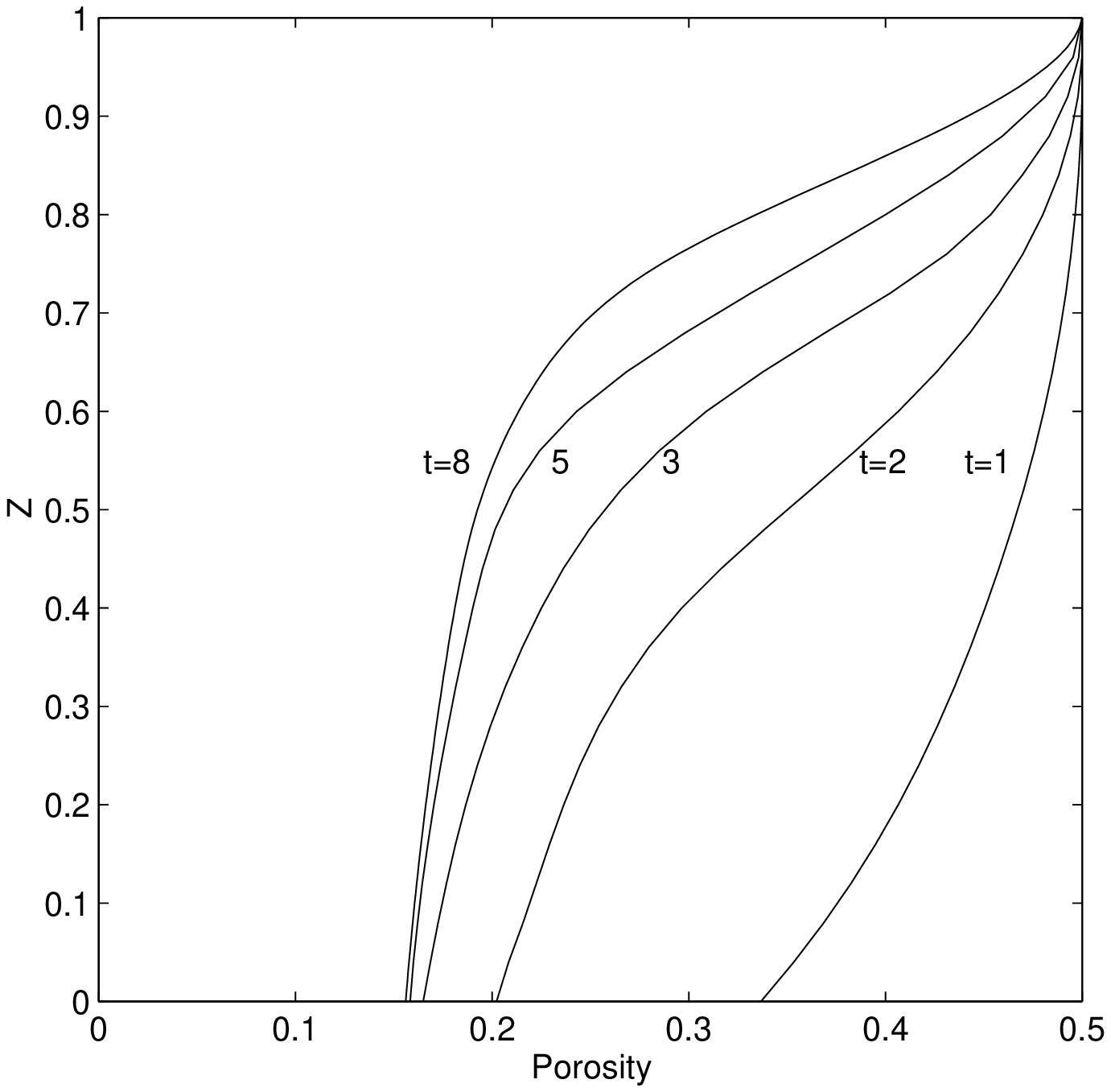}{Viscous compaction profile of porosity versus
          the rescaled height $Z$. All the  other  parameters are the
          same as in  Figure 1. A viscous compaction relation
          between effective pressure and velocity is used.
          The profile now is nearly parabolic.   }

\fig{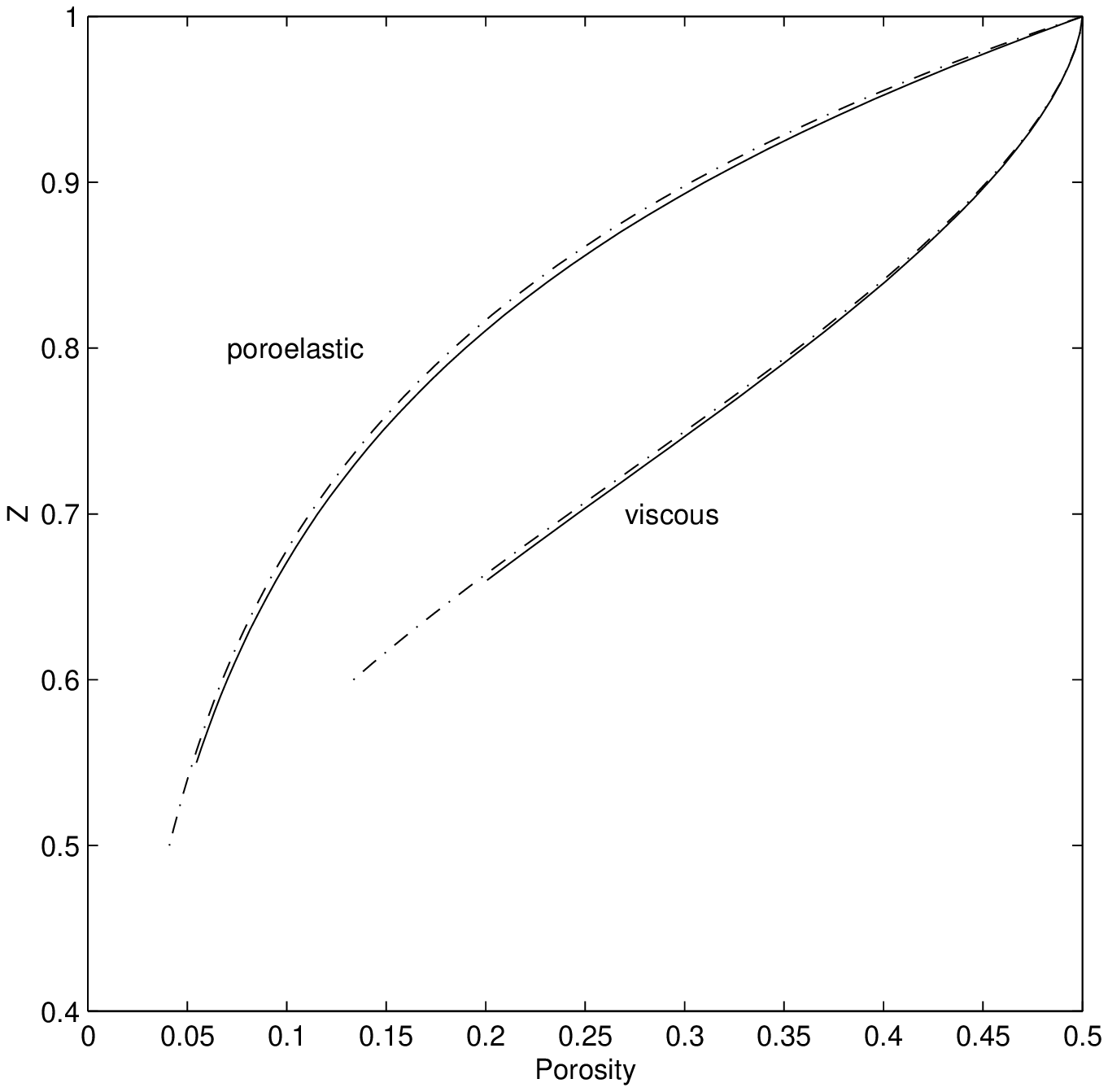}{Comparison of asymptotic solutions (\ref{equ-500})
          and (\ref{equ-500-500}) (dashed curves) with  numerical results
          (solid curves) in the
          top region where the profile is nearly at equilibrium state.
          The agreement is    clearly shown.}

\end{document}